\documentclass[conference]{IEEEtran}
\IEEEoverridecommandlockouts
\usepackage{cite}
\usepackage{amsmath,amssymb,amsfonts}
\usepackage{algorithmic}
\usepackage{graphicx}
\usepackage{textcomp}
\usepackage{xcolor}
\usepackage{comment}

\def\BibTeX{{\rm B\kern-.05em{\sc i\kern-.025em b}\kern-.08em
    T\kern-.1667em\lower.7ex\hbox{E}\kern-.125emX}}

\makeatletter
 \let\old@ps@headings\ps@headings
 \let\old@ps@IEEEtitlepagestyle\ps@IEEEtitlepagestyle
 \def\confheader#1{%
 \def\ps@headings{%
 \old@ps@headings%
 \def\@oddhead{\strut\hfill#1\hfill\strut}%
 \def\@evenhead{\strut\hfill#1\hfill\strut}%
 }%
 \def\ps@IEEEtitlepagestyle{%
 \old@ps@IEEEtitlepagestyle%
 \def\@oddhead{\strut\hfill#1\hfill\strut}%
 \def\@evenhead{\strut\hfill#1\hfill\strut}%
 }%
 \ps@headings%
 }
 \makeatother

\confheader{%
2023 IEEE International Conference on Computing (ICOCO)
 }

\begin{document}

\title{Survey on Security Attacks in Connected and Autonomous Vehicular Systems}

\author{

\IEEEauthorblockN{S M Mostaq Hossain}%
\textit{IEEE Student Member}\\
\IEEEauthorblockA{\textit{Dept. of Computer Science} \\
\textit{Tennessee Tech University}\\
Tennessee, USA \\shossain42@tntech.edu}

\and 

\IEEEauthorblockN{Shampa Banik}%
\textit{IEEE Student Member}\\
\IEEEauthorblockA{\textit{Dept. of Computer Science} \\
\textit{Tennessee Tech University}\\
Tennessee, USA \\
sbanik42@tntech.edu}

\and
   
\IEEEauthorblockN{Trapa Banik}%
\textit{IEEE Student Member}\\
\IEEEauthorblockA{\textit{Dept. of ECE}\\
\textit{Tennessee Tech University}\\
Tennessee, USA \\tbanik42@tntech.edu}

\and 

\IEEEauthorblockN{Ashfak Md Shibli}%
\textit{IEEE Student Member}\\
\IEEEauthorblockA{\textit{Dept. of Computer Science} \\
\textit{Tennessee Tech University}\\
Tennessee, USA \\
ashibli42@tntech.edu}

}

\maketitle

\begin{abstract}
Connected and autonomous vehicles, also known as CAVs, are a general trend in the evolution of the automotive industry that can be utilized to make transportation safer, improve the number of mobility options available, user costs will go down and new jobs will be created. However, as our society grows more automated and networked, criminal actors will have additional opportunities to conduct a variety of attacks, putting CAV security in danger. By providing a brief review of the state of cyber-security in the CAVs environment, this study aims to draw attention to the issues and concerns associated with security. The first thing it does is categorize the multiple cyber-security threats and weaknesses in the context of CAVs into three groups: attacks on the vehicle's network, attacks on the Internet at large, and other attacks. This is done in accordance with the various communication networks and targets under attack. Next, it considers the possibility of cyber attacks to be an additional form of threat posed by the environment of CAVs. After that, it details the most up-to-date defense tactics for securing CAVs and analyzes how effective they are. In addition, it draws some conclusions about the various cyber-security and safety requirements of CAVs that are now available, which is beneficial for the use of CAVs in the real world. At the end, we discussed some implications on Adversary Attacks on Autonomous Vehicles. In conclusion, a number of difficulties and unsolved issues for future research are analyzed and explored.
\end{abstract}


\section{Introduction}
Improvements in automotive technology have accelerated during the past decade. The traditional idea of safety in the automobile industry has been revolutionized as a result of the increasing complexity of modern vehicle systems, which coincides with a rapid rise in the use of electronic components and wireless technologies. In addition, there has been a rise in the interest in the development of vehicular networks (VN)\cite{b1} and connected autonomous vehicles (CAV)\cite{b2}, which has led to the emergence of new security issues and vulnerabilities. In the meanwhile, the industry standards for in-vehicle and vehicular communications do not adhere to the well-established computer security regulations due to the hardware limits and changes in network architecture. The market for CAVs is rapidly expanding around the world. The forecast for 2050 puts it at \$7 trillion \cite{b3}. In addition, numerous major automakers and IT behemoths are working furiously to beat the clock and deliver usable CAVs to the market. Several ambitious research initiatives have been launched around the world due to the substantial business potential of CAVs.

CAVs could revolutionize many industries, but they also present serious privacy and safety concerns \cite{b4}. Specifically, the rising degree of interconnectivity and mechanization raises the stakes for security threats. An attacker might take control of a vehicle or its accessories. Information sharing and wireless connectivity are primary drivers of successful cyber attacks on automobiles. Consequently, some of the most pressing security and privacy concerns for IVs are related to the confidentiality of user identity and data, the integrity of data transmissions (both incoming and outgoing), and the safety of the vehicle's Electronic Control Units (ECUs) \cite{b5}.

In this paper, we talk about how we think about the security of connected vehicles based on recent threats and ways to protect against them. By giving an overview of these threats, we show how important it is that connected vehicles have good security, in general, to protect them from cyber-attacks. We also show that secure systems would be broken even if one of the security features wasn't covered by any of the three categories. We think this paper is helpful for making and using connected vehicle services, as well as researching cyber security for connected vehicles.

This paper's remaining sections are structured as follows. In Section II, we present a literature review on the topic of CAV cyber-security. In Section III, we present the implications of Adversary Attacks on Autonomous Vehicles, and in Section IV we discuss the limitations. Section V is the final section of this work.

\begin{figure*}
  \includegraphics[width=\textwidth,height=11cm]{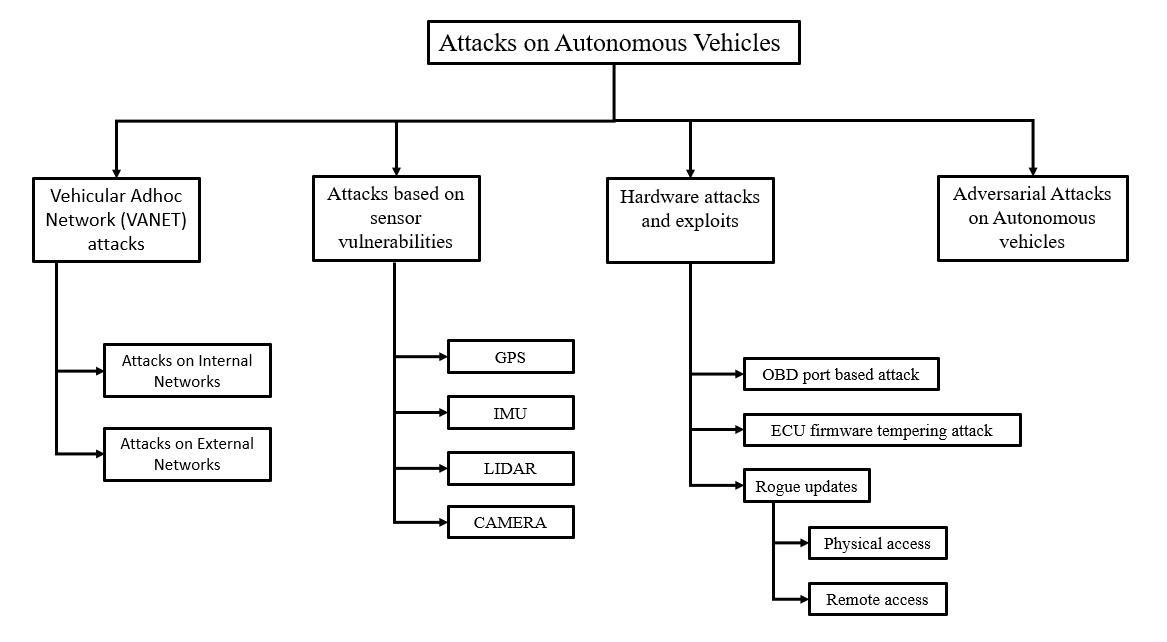}
  \caption{The potential attacks in connected and autonomous vehicles.}
\end{figure*}

\section{Research Overview}
Autonomous vehicles, in contrast to conventional ones, must communicate with external networks consisting of other vehicles and infrastructure. These external networks present both a potential attack vector and a hacking opportunity \cite{b6}. Autonomous vehicles rely on the processing of enormous data sets to function. When dealing with image data, deep learning algorithms are especially susceptible to adversarial attacks, and the likelihood of false positives increases with the volume of data to be processed. Attackers can take advantage of flaws and vulnerabilities in autonomous vehicles \cite{b7}, and the technology is still in its infancy in terms of both hardware and software, making it challenging to make it dependable under all scenarios.

\subsection{Attacks on Vehicle-Based Ad-Hoc Networks (VANETs)}
\begin{enumerate}
\item Internal Network Attacks
\begin{enumerate}
    \item key and Passcode attacks\\
    One of the safeguards of linked cars is a passcode and keys. Passwords and keys using IR-based technology may be cracked after multiple failed attempts. The passcodes of linked vehicles can be broken by a brute-force attack \cite{b8}. Passengers' personal information may be compromised by a disruption in Bluetooth connectivity.
\end{enumerate}
\item External Network Attacks
\begin{enumerate}
    \item V2X Attacks on Networks\\
        There are numerous advantages to the idea of linked automobiles, and new technology in vehicular networks makes it possible for vehicles to communicate data with one another and with fixed infrastructure, but there is also a risk that cybercriminals would exploit this to get access to the underlying network and commit crimes \cite{b9}. Its capacity to link to a smartphone, the cloud, and other devices is described by its V2X communication. Dedicated Short Range Communication (DSRC) is a duplex communication protocol channel designed for automotive use that operates at 5.9 GHz with a bandwidth of 75 Mhz, while IEEE 802.11 p and WAVE are commonly used for vehicle-to-everything (V2X) communication. They are all susceptible to attack because of recognized flaws.
        Overtaking, changing lanes, and exchanging information with oncoming cars all rely on data exchanged over vehicle-to-vehicle (V2V) networks, which might be vulnerable to attacks. An impersonation attack involves a hostile car connecting to a host vehicle using a spoofed identifier, establishing contact with the host vehicle, and then sending and receiving malicious and sensitive data, which is then logged and stored[8].
        Unencrypted and insecure protocols are a fundamental flaw of V2V communication since they allow attackers to listen in on conversations between vehicles and steal important information like authentication keys, which can subsequently be used in authentication attacks.
        Threats to Vehicle-to-Infrastructure (V2I) Communications: When a vehicle establishes a connection with the base station, it opens up a two-way line of communication. Through the car's connection to vulnerable intelligent traffic signs and cellular network nodes, attackers can get unauthorized access to the network and ECUs of the vehicle.
        \item Distributed Denial Of Services (DDOS)\\
        In the case of Internet-connected vehicles, denial-of-service attacks are among the most dangerous possibilities. system service is denied as a result of several attack mechanisms, causing congestion and infrastructure damage \cite{b10}. As a result, there is a risk that the vehicles would collide, which might be fatal for the people inside.
    \end{enumerate}
\end{enumerate}

\begin{table*}[]
\caption{Summary of various attacks on Connected and Autonomous Vehicular Systems}
\label{tab:my-table}
\begin{tabular}{|l|l|l|l|l|l|}
\hline
\textbf{Sl.} & \textbf{Types} & \textbf{Impacts} & \textbf{Attack methods} & \textbf{Features} & \textbf{Year {[}Ref.{]}} \\ 

\hline

\multicolumn{1}{|p{0.4em}}{1} &    
\multicolumn{1}{|p{10em}|}{Remote sensor attacks} &
\multicolumn{1}{p{13em}|}{Regard injected signals as legal signals} & 
\multicolumn{1}{p{13em}|}{Various attack vectors }& 
\multicolumn{1}{p{13em}|}{Cannot distinguish the legal input and deceived} &
\multicolumn{1}{p{4em}|}{2022\cite{b23}}\\

\hline

\multicolumn{1}{|p{0.4em}}{2} &    
\multicolumn{1}{|p{10em}|}{GPS spoofing attacks} &
\multicolumn{1}{p{13em}|}{Manipulate the received GPS signals arbitrarily} & 
\multicolumn{1}{p{13em}|}{Broadcast fake GPS signals with high signal strength}& 
\multicolumn{1}{p{13em}|}{Ignore true GPS signals} &
\multicolumn{1}{p{4em}|}{2021\cite{b24}}\\

\hline

\multicolumn{1}{|p{0.4em}}{3} &    
\multicolumn{1}{|p{10em}|}{Location trailing attacks} &
\multicolumn{1}{p{13em}|}{Discover behaviors and activities of the vehicle, the profile of the driver} & 
\multicolumn{1}{p{13em}|}{Relate pseudonymous position samples to specific vehicles}& 
\multicolumn{1}{p{13em}|}{A single pseudonym is not sufficient to protect vehicles' privacy} &
\multicolumn{1}{p{4em}|}{2022\cite{b25}}\\

\hline

\multicolumn{1}{|p{0.4em}}{4} &    
\multicolumn{1}{|p{10em}|}{Close proximity vulnerabilities} &
\multicolumn{1}{p{13em}|}{The system or communication mechanism may be attacked} & 
\multicolumn{1}{p{13em}|}{Bluetooth or key-less entry and ignition systems}& 
\multicolumn{1}{p{13em}|}{The attack may be primitive and happen occasionally} &
\multicolumn{1}{p{4em}|}{2021\cite{b26}}\\

\hline

\multicolumn{1}{|p{0.4em}}{5} &    
\multicolumn{1}{|p{10em}|}{CAN and SAE J1939 buses vulnerabilities} &
\multicolumn{1}{p{13em}|}{Unauthorized nodes can eavesdrop on any conversation taking place on the bus} & 
\multicolumn{1}{p{13em}|}{The result of the system of arbitration}& 
\multicolumn{1}{p{13em}|}{An attacker can compromise a network by inserting a malicious frame} &
\multicolumn{1}{p{4em}|}{2020\cite{b27}}\\

\hline

\multicolumn{1}{|p{0.4em}}{6} &    
\multicolumn{1}{|p{10em}|}{Attacks using a flashing ECU software} &
\multicolumn{1}{p{13em}|}{Disclose information, corrupt information, degrade hardware performance, trick a remote device} & 
\multicolumn{1}{p{13em}|}{Reverse engineering, code modification, fuzzing attacks}& 
\multicolumn{1}{p{13em}|}{Use unpatched vulnerabilities to gain access} &
\multicolumn{1}{p{4em}|}{2021\cite{b28}}\\

\hline

\multicolumn{1}{|p{0.4em}}{7} &    
\multicolumn{1}{|p{10em}|}{Business process integration scams} &
\multicolumn{1}{p{13em}|}{Integrated business services attacks} & 
\multicolumn{1}{p{13em}|}{Gain the client-level access, system-level access, access to the other vehicle systems}& 
\multicolumn{1}{p{13em}|}{The adversary uses the client-level access to implement another attack} &
\multicolumn{1}{p{4em}|}{2021\cite{b26}}\\

\hline

\multicolumn{1}{|p{0.4em}}{8} &    
\multicolumn{1}{|p{10em}|}{DoS attacks} &
\multicolumn{1}{p{13em}|}{Delay and interfere the receiver's response} & 
\multicolumn{1}{p{13em}|}{Insert useless messages, create some issues on network nodes}& 
\multicolumn{1}{p{13em}|}{Dangerous and fatal} &
\multicolumn{1}{p{4em}|}{2023\cite{b29}}\\

\hline

\multicolumn{1}{|p{0.4em}}{9} &    
\multicolumn{1}{|p{10em}|}{Impersonation attacks} &
\multicolumn{1}{p{13em}|}{Send phony messages, modify received messages, and discard important ones} & 
\multicolumn{1}{p{13em}|}{Fake or use another identity}& 
\multicolumn{1}{p{13em}|}{Multiple identities are spoofed} &
\multicolumn{1}{p{4em}|}{2021\cite{b30}}\\

\hline

\multicolumn{1}{|p{0.4em}}{10} &    
\multicolumn{1}{|p{10em}|}{Replay attacks} &
\multicolumn{1}{p{13em}|}{Confound authorities, mislead traffic, endanger transportation} & 
\multicolumn{1}{p{13em}|}{Record and re-transmit the early valid packets later}& 
\multicolumn{1}{p{13em}|}{Occur in certain procedures for securing and exchanging keys} &
\multicolumn{1}{p{4em}|}{2020\cite{b31}}\\

\hline

\multicolumn{1}{|p{0.4em}}{11} &    
\multicolumn{1}{|p{10em}|}{Routing attacks} &
\multicolumn{1}{p{13em}|}{Disturb the normal routing process, drop passing packets} & 
\multicolumn{1}{p{13em}|}{One compromised or cooperative node}& 
\multicolumn{1}{p{13em}|}{Exploit the drawback and vulnerability of routing protocols} &
\multicolumn{1}{p{4em}|}{2020\cite{b32}}\\

\hline

\multicolumn{1}{|p{0.4em}}{12} &    
\multicolumn{1}{|p{10em}|}{Data falsification attacks} &
\multicolumn{1}{p{13em}|}{Mislead receivers' reaction and result in fatalities} & 
\multicolumn{1}{p{13em}|}{Distribute or disseminate fabricated security alerts and other information}& 
\multicolumn{1}{p{13em}|}{An unnecessary warning can open the way ahead. Congestion in the event of a falsified accident message} &
\multicolumn{1}{p{4em}|}{2022\cite{b33}}\\

\hline

\multicolumn{1}{|p{0.4em}}{13} &    
\multicolumn{1}{|p{10em}|}{Eavesdropping attacks} &
\multicolumn{1}{p{13em}|}{Leak actual identity and private key} & 
\multicolumn{1}{p{13em}|}{Overhear the wireless medium}& 
\multicolumn{1}{p{13em}|}{The victim is not conscious of the attacks} &
\multicolumn{1}{p{4em}|}{2021\cite{b34}}\\

\hline

\multicolumn{1}{|p{0.4em}}{14} &    
\multicolumn{1}{|p{10em}|}{Password and key attacks} &
\multicolumn{1}{p{13em}|}{Crack password or key} & 
\multicolumn{1}{p{13em}|}{Multiple values are examined until the mechanism is compromised}& 
\multicolumn{1}{p{13em}|}{The attacks fall into three groups} &
\multicolumn{1}{p{4em}|}{2021\cite{b26}}\\

\hline

\multicolumn{1}{|p{0.4em}}{15} &    
\multicolumn{1}{|p{10em}|}{Infrastructure attacks} &
\multicolumn{1}{p{13em}|}{Cut off the communication in the infrastructure} & 
\multicolumn{1}{p{13em}|}{Vulnerability related to message transfer}& 
\multicolumn{1}{p{13em}|}{Attacks are related to cloud service} &
\multicolumn{1}{p{4em}|}{2020\cite{b35}}\\

\hline

\multicolumn{1}{|p{0.4em}}{16} &    
\multicolumn{1}{|p{10em}|}{Conditional attacks} &
\multicolumn{1}{p{13em}|}{Create unsafe environments and critical safety issues} & 
\multicolumn{1}{p{13em}|}{N/A}& 
\multicolumn{1}{p{13em}|}{Communicated data differ arbitrarily from reality. No deviations exceed the threshold} &
\multicolumn{1}{p{4em}|}{2021\cite{b26}}\\

\hline

\multicolumn{1}{|p{0.4em}}{17} &    
\multicolumn{1}{|p{10em}|}{Intrusions into AI defenses} &
\multicolumn{1}{p{13em}|}{Modify training algorithm and training data} & 
\multicolumn{1}{p{13em}|}{Data poisoning, escape, theft, and inference attacks on models}& 
\multicolumn{1}{p{13em}|}{Take place throughout the data-gathering, data-training, and data-predicting stages} &
\multicolumn{1}{p{4em}|}{2020\cite{b36}}\\

\hline

\end{tabular}
\end{table*}

\subsection{Attacks based on sensor vulnerabilities}
\begin{enumerate}
    \item Global Positioning System (GPS)\\
    Using GPS data with a high degree of accuracy allows one to both identify and navigate the car. To get around the difficulties of accessing GPS data, the number of satellites in the public domain was increased so that anyone with internet access could simply acquire the data. Data transparency and open access make it easier for hackers to manipulate information and provide false directions or even take over the vehicle's navigation system \cite{b11}. Because of this, there were concerns raised about the passengers' safety and security. GPS spoofing and jamming are two terms that refer to the same activity, which involves hackers transmitting data and signals that are not accurate to the GPS system. As the GPS sensors are adjusted to pick up greater signals, the unrealistic signal grows stronger and the vehicle's location drifts further and further from the target \cite{b12}.
    \item Inertial Measurement Unit (IMU)\\
    IMU stands for inertial measurement unit and is comprised of a gyroscope and an accelerometer. Its purpose is to measure the vehicle's speed, as well as its acceleration and orientation. They also keep an eye on the ways in which environmental dynamics, such as the gradient, are shifting. The data produced by the sensors can be adjusted or left unchanged in order to determine the road's incline. This causes the car to travel more slowly on inclines, which in turn slows the traffic behind it \cite{b13}.
    \item Light Detection and Ranging (Lidar)\\
    The technique of Light Detection and Ranging (LiDAR) can be utilized for the purposes of localization, obstacle detection, and collision avoidance. The system relies on the time it takes for the light to travel to and from the vehicle to determine the distance to the object. In the event that the hacker transmits a signal to the scanner at the same frequency, it is safe to presume that the object has been located \cite{b14}. As a result, the autonomous vehicle will either proceed more slowly or come to a complete halt.
    \item Monoscopic and Stereoscopic Cameras\\
    Cameras can perform a wide variety of detecting tasks, including lane detection, traffic sign recognition, headlight detection, obstacle identification, and more. It is possible to disable the functionality of cameras by shining high-beam torches or the headlights of the opposing vehicle in their direction. It is possible that it will give rise to safety problems such as false detection or the failure to identify the items. The camera's complementary metal oxide (CMOS) sensors can be damaged by extremely bright illumination \cite{b15}.
\end{enumerate}
\subsection{Exploits and attacks against hardware}
\begin{enumerate}
    \item Attack based on OBD port\\
    Most cars manufactured after 2008 will include what is called an "OBD" port, which stands for "onboard diagnostics \cite{b9}. The onboard diagnostics (OBD) port is utilized to acquire vehicle diagnostic information. A vehicle's performance and any problems it may have are logged and shown. CAN bus is used for communication between electronic control units. It's a USB-like handheld device that plugs into a port often found below the dashboard, across from the driver's seat. which can then be connected to a computer through a USB port or wirelessly by Bluetooth \cite{b16}. A malicious actor could compromise a car's network by communicating with the vehicle's ECUs through a PC connected to the vehicle.

    \item Threat of Firmware Tampering to ECU\\
    More than one hundred electronic control modules called engine control units (ECUs) regulate the sensors and actuators of every vehicle subsystem \cite{b17}. Despite the fact that ECU code is propitiatory and therefore secure, attackers have started to re-flash them with malicious custom firmware in recent attacks. Since we suppose the attacker has actual access to the ECU, we label this as a direct access attack. The external interface is used by the attacker to upgrade the firmware of the ECU, changing the ECU's behavior. Using authentication for software updates and hashing techniques to ensure the security of the ECU firmware code during its lifetime \cite{b17}; modifying the ECU memory; tampering with security keys.
    \item Rogue updates\\
    One source of Rogue upgrades is the latest firmware in connected vehicles. These patches aren't officially sanctioned by the manufacturer, and they're missing critical security and safety patches. Vehicle data is vulnerable to serious cyber attacks that can reveal personal information. To the extent that sufficient security flaws are provided, hackers can exploit them to plant malware in the connected cars' firmware and take command of them. Exploitation may occur via: Physical Access or remote access
    \begin{enumerate}
    \item Physical Access\\
    Due to the increasing direct integration of the physical layer with the ECUs, the danger of cyber-attacks has increased in recent years \cite{b17}. All the modules, including the sensors, controls, and communications, are vulnerable to hacking. Overloading the vehicle's physical layer or attacking individual electrical modules are two examples of possible attacks.
    \item Remote Access\\
    Various networks, including WiFi, Bluetooth, 4G, etc., provide remote access. Unintentional direct connection of ECUs to the CAN bus \cite{b17}. Because of its internet connection, the firmware could be compromised by malicious code. The automotive industry is uncertain about the nature of the dangers they face or the best course of action to eliminate them. Even the automakers themselves are unsure of the best course of action for recovery.
    \end{enumerate}
\end{enumerate}

\subsection{Autonomous vehicle adversarial attacks}
Probabilistic estimations provide the basis of Deep Learning's ability to perform categorization tasks. When the likelihood that a set of data is associated with a particular class is high, we say that confidence in that class is high \cite{b18}. Autonomous vehicles utilize cutting-edge sensing and perception technology, and deep learning algorithms like Deep Neural Networks (DNN) \cite{b19}. However, DNNs can be breached by specific malicious attacks, making it possible to launch such attacks on autonomous cars. Adversarial attacks are a form of cyberattack in which the target DNN is tricked or misled into making potentially harmful judgments by slight modifications or additions to the original data \cite{b20}. In \cite{b21}, stop signs can be modified physically by applying stickers to specific areas. It was misread by the software designed to identify traffic signs.

\subsection{Miscellaneous Attacks}
Many researchers and hackers were hard at work putting together a variety of different approaches. The keyless entry system is vulnerable to spoofing, malware attacks, and programs installed in the infotainment and navigation systems. New varieties of attacks, such as those that target electric vehicles by taking advantage of the electric charger, which connects the devices and logs the data. The variety of OEMs and tier-1 suppliers in the automotive industry increases the likelihood of a wide variety of vulnerabilities and attacks. This is because of the many different types of cars available, each with its own advantages and disadvantages, as well as the advent of autonomous driving technology. This complicates efforts to ensure the safety and visibility of several original equipment manufacturers.
In Table I we have tried to draw a brief picture of different types of attacks on connected autonomous vehicle systems. In addition, digital twin technology \cite{b37} is about to be adopted by the automotive sector. It's a new frontier in the study of autonomous, linked vehicles. Although the digital twin offers superior security defense, there have been certain difficulties from an attack standpoint. As a result, there has been a progressive shift in the scope of the study. 

\section{Research problem definition}
The Adversary Attacks on Autonomous Vehicles could be Studied in Detail. With the help of the concept of the author's suggestion \cite{b22}, the following three implications could be the solution for that.
\begin{enumerate}
    \item \textbf{First, it's crucial to use a variety of defensive strategies in concert.}\\
    There is no silver bullet for protecting driving models from the five different ways they might be attacked. While feature squeezing has a good attack detection rate for all attacks, it has a high false positive rate of 40\% for regular input pictures. It is recommended to use a combination of these defensive measures to ward off enemies. It's possible that automobile sides aren't the best place for detection techniques that require a lot of processing power. Edge computing allows for a quicker response time, and the addition of hostile detection middleware on edge nodes can increase the security of autonomous vehicles \cite{b22}.
    \item \textbf{Second, more research is required to understand how various DNN topologies of regression models affect their susceptibility.}\\
    VGG16 is safer than previous versions in both white-box and black-box attacks. The complex model structure of VGG16 makes it more challenging to solve. It is possible that more complicated models are more robust in terms of safety. Due to computational constraints, autonomous cars can't transport highly complicated driving models; therefore, we need to create a model with just the proper level of structure complexity and processing requirements. With the help of edging computing, even complex driving models may be put into use. Complex models can be abstracted and communicated to vehicles, roadside units, and the cloud in order to reduce the strain on the vehicle's resources\cite{b22}.
    \item \textbf{Third, Protecting driving model detail and researching adversarial example transferability is critical.}\\
    The results of the white-box assault experiment (RQ1) reveal that driving models are vulnerable to attacks and that attackers who have knowledge of their attributes may be able to devise effective adversarial attacks. Hide hyper-parameters and driving model neural network structure. Information extraction can be used to attack deep learning algorithms, according to a recent study. Exploring methods like preventing model extraction. The black box attack experiment (RQ2) shows that hostile cases generated on one driving model aren't transferrable. This contradicts classification model attack transferability findings. Driving models are regression models, hence transferability differences between classification and regression models should be studied. Different regression methods can learn V-B hyper-planes. Test this hypothesis. Driving model transferability research will help construct and defend against black-box attacks\cite{b22}.
\end{enumerate}
\section{Limitations}
The work \cite{b22} suggests that hackers may be able to infiltrate a self-driving car's system during an over-the-air software or firmware update. Malware could intercept photos and produce hostile samples before perception. Hostile instances cannot be seen by humans or detected by the car's infotainment system, unlike direct interference with vehicle components. In contrast to adversarial attacks, basic attacks, such as replacing the driving scenery with random pictures, are easy to identify. We tested three different CNN-based driving models. Due to their reliance on sequence-to-sequence structures and input driving videos, the top three driving models on Udacity's leaderboard were not selected. Untested CNN with RNN models exist. The Udacity dataset is used only in our experiments, as before. More primary models and data sets are required to further our findings. CleverHans and Foolbox are open-source adversarial attack and defense software. Three classifier tools. Steering angle regression models will require adapting these tools. Avoid feature squeezing with optimization. This technique generates one valid adversarial example on MNIST in 2030 seconds. We didn't use it because it's too slow for real-time driving model attacks.

\section{Conclusion}
In this paper, In this document, we've tried to include most of the most common cyber attacks and exploits that can be used against a connected, autonomous vehicle. This is the hottest area, and attacks in it are constantly evolving in sophistication. Meanwhile, hackers are always coming up with new ways to trick and hack the vehicle. However, the security component of vehicles is not getting the attention that it deserves as the development of autonomous vehicles continues to proceed at a rapid pace. At a time when many nations are making efforts to have autonomous vehicles on the road as quickly as feasible, this could pose a serious threat to the adoption and security of autonomous vehicles. Scientists should stop squabbling and start working together on projects that will finally give cyber security the attention it deserves during the planning and development phases. Finally, we address Adversary Attacks on Autonomous Vehicles and offer some consequences based on our findings.

\end{document}